\documentclass[twocolumn,superscriptaddress]{revtex4}
\usepackage{graphicx}
\usepackage{amsmath}
\usepackage{amsfonts}
\usepackage{amssymb}
\usepackage{color}
\usepackage{amscd}
\usepackage{bm}

\begin{document}
\title{Running of effective dimension and cosmological entropy in early universe}
\author{Yong Xiao}
\email{xiaoyong@hbu.edu.cn}
\affiliation{Key Laboratory of High-precision Computation and Application of Quantum Field Theory of Hebei Province,
College of Physical Science and Technology, Hebei University, Baoding 071002, China}
\affiliation{Department of Physics and Astronomy, University of Sussex, Brighton, BN1 9QH, United Kingdom}

\begin{abstract}
In this paper, we suggest that the early universe starts from a high-energetic state with a two dimensional description and the state recovers to be four dimensional when the universe evolves into the radiation dominated phase. This scenario is consistent with the
recent viewpoint that quantum gravity should be effectively two dimensional in the ultraviolet
and recovers to be four dimensional in the infrared. A relationship has been established between the running of effective dimension
and that of the entropy inside particle horizon of the universe, i.e., as the effective dimension runs from two to four,
the corresponding entropy runs from the holographic entropy to the normal entropy appropriate to radiation.
These results can be generalized to higher dimensional cases.
\end{abstract}
 \maketitle

\section{Introduction}
 It has been established that black holes have thermodynamic properties such as temperature and entropy. In particular, the thermodynamics of the Schwarzschild black hole of radius $R$ has the form
\begin{align}
M = \frac{R}{2G},\ \ \  T = \frac{1}{4\pi R},\ \ \ S = \frac{A}{4G},\label{thermo}
\end{align}
where we have set $\hbar=c=K_B=1$. The celebrated Bekenstein--Hawking entropy $\frac{A}{4G}$ is often called holographic entropy because of its proportionality with the boundary area of the system. In \cite{bema} Bekenstein and Mayo revealed a secret behind this kind of thermodynamics, that is, black holes are effectively $1+1$ dimensional as far as entropy flow is concerned. Recently, in \cite{xiao} Xiao showed more directly that the thermodynamics \eqref{thermo} is $1+1$ dimensional in essence. Quantum gravitational (QG) particles with non-trivial phase space have been introduced there in order to provide a microscopic explanation to \eqref{thermo}, with the equation of state (EoS) $w\equiv P/\rho=1$ derived as a byproduct.

In the context of cosmology, we mainly concern about the dominate stage of these QG particles in the evolvement history of the universe. First, we expect the very-early high-energetic stage of the universe be controlled by QG theory. Second, by Friedmann equations the evolvement of the universe declines to lower the value of $w$ as time increases. So it is natural to expect an early stage of the universe with $w=1$ exists before the radiation dominated universe with $w=1/3$. It immediately follows an interesting scenario of the evolution of the very early universe: When the dominate EoS evolves from $w=1$ to $w=1/3$, the effective description of the universe runs from $1+1$ dimensional to $3+1$ dimensional, along with a remarkable evolvement of the entropy from holographic entropy for QG particles to normal entropy for radiation.

Interestingly, in recent years, there have been cumulative evidences \cite{laus,amb,horava} indicating that quantum gravity should be effectively $1+1$ dimensional at small sizes (commonly near the Planck length $l_p$ but not necessarily transcending it \footnote{There are even attempts to lower the characteristic scale of dimensional reduction down to near the Tev scales. See \cite{carlip1} for examples.}) and recovers to be $3+1$ dimensional at large scales. Actually the phenomenon of short-distance dimensional reduction was obtained from various approaches to quantum gravity and various definition of effective dimensions. The universality has even been viewed as a curious question to be addressed \cite{carlip1,carlip2}. The subtlety here is the concept of ``effective dimension" which is defined and obtained by examining some specific physical behaviors sensitive to space-time dimensions. For example, the diffusive behavior of particles defines the spectral dimension, and the temperature dependence of the thermodynamic quantities determines the thermodynamic dimension. In particular, Ho\v{r}ava-Lifshitz gravity flows to be $2$ dimensional in the ultraviolet, measured by both the generalized spectral dimension and thermodynamic dimension \cite{horava}. Ho\v{r}ava stressed that the behavior doesn't necessarily imply a foamy structure of space-time and can be explained by the existence of some special properties of quantum gravity such as the anisotropic scaling of space and time at short distances, keeping the $3+1$ dimensional geometry smooth. Thus to avoid confusion, it would be better to distinguish clearly between the two concepts of effective dimension and space-time dimension: The effective dimension coincides with the given space-time dimension $D$ only at large scales, but deviates from it and reduce to $2$ at small scales. We refer the interested reader to \cite{carlip1,carlip2} for a clear discussion of the concept of effective dimensions and the phenomenon of short-distance dimensional reduction in quantum gravitational theories.

The paper is organized as follows. First we review the thermodynamics \eqref{thermo} can be microscopically explained by introducing the so-called QG particles and it is effectively $1+1$ dimensional \cite{xiao}. Next we study the universe filled with a single kind of constituents for simplicity and derive the concerned properties including the entropy inside particle horizon and effective dimension. Then we provide a self-consistent evolvement scenario for the early universe, with emphasis on the running of effective dimension and entropy. We always use the thermodynamic dimension as the dimensional estimator for studying the phenomenon of dimensional reduction, which is also a good one among others \cite{carlip1,carlip2}. We exhibit the general results in $D=d+1$ dimensions in the Appendix.

\section{QG particles and Bekenstein-Hawking entropy}
The fact that black holes have temperature and entropy implies that there must be some kinds of microscopic degrees of freedom behind it. In order to provide a statistical interpretation to the thermodynamic behaviors \eqref{thermo}, we consider the Schwarzschild black hole as composed of microscopic particles that we call QG particles for convenience \cite{xiao}. In fact such ideas are not new, for example, the charged AdS black holes have been suggested to be consisting of ``molecules'' with attractive or repulsive interactions \cite{wei,wei2}.

We take the QG particles as massless bosonic particles with the logarithm of partition function written as \footnote{The form of \eqref{thermo} is so simple that we should avoid to introduce all unnecessary parameters such as particle mass, particle number and interaction strength.}
\begin{align}
\begin{split}
\ln \Xi = - \sum\limits_{i} \ln ( 1 - e^{ - \beta \varepsilon_i} )= - g \int_0^\infty  \ln ( 1 - e^{ - \beta \varepsilon }  ) D(\varepsilon) d\varepsilon,
 \end{split} \label{lnx}
\end{align}
where $\beta=1/T$, $D(\varepsilon)d\varepsilon$ is the number of quantum states with energy between $\varepsilon$ and $\varepsilon+d\varepsilon$, and $g$ represents other possible degrees of freedom such as polarization.

 Before handling with the black hole thermodynamics, we look back to the familiar statistical mechanics of photon gas system for comparison. Quantum mechanics and special relativity respectively tells that $\triangle q_i \triangle p_i \geq \frac{\hbar}{2}$ and $\varepsilon=c p$ (we only restore the fundamental constants temporarily). Then the quantum states of a photon can be labeled by $\vec{p}=\frac{2\pi \hbar}{L}(m_1,m_2,m_3)$ with energy spectrum $\varepsilon=c|\vec{p}|$. So the number of quantum states between $\varepsilon$ and $\varepsilon+d\varepsilon$ can be evaluated by $D(\varepsilon) d\varepsilon= \frac{1}{2 \pi^2 } V \varepsilon ^2 d\varepsilon$. Substituting it and $g=2$ into eq.\eqref{lnx}, there is $\ln \Xi =\frac{ \pi^2}{45  } \frac{V}{\beta^3}$. The standard photon gas thermodynamics follows
\begin{align}
E = - \frac{\partial }{\partial \beta}  \ln \Xi =\frac{\pi^2  }{15  }V T^4, \label{QFTe} \\
S = k_B (\ln \Xi  + \beta E ) =\frac{4 \pi^2 }{45 } V T^3, \label{QFTs}\\
P =T \frac{\partial \ln \Xi }{\partial V}  = \frac{1}{3}\rho.
\end{align}
When the energy is of the same order of that of a black hole of the same size, $E\sim E_{bh}\sim R/G$, there is $T_{max} \sim L^{-1/2}$. Substituting it into eq.\eqref{QFTs}, we get the entropy bound for conventional quantum field theory (QFT) \cite{hooft,bousso,cohen,us1}
\begin{align}
S_{QFT}   \sim A^{3/4}.
\end{align}

Now turn to the system consisting of QG particles. According to \cite{xiao}, in order to account for the thermodynamics \eqref{thermo}, we should take
\begin{align}
g D(\varepsilon) d\varepsilon=\frac{ 9 V}{\pi G }d\varepsilon.  \label{state2}
\end{align}
The special form \eqref{state2} means that QG particles have a distinctive energy spectrum from that of photons, which may hint some radical modification to the basic physical principles. Whatever, in practice we can consider all the non-trivial physical effects have been encapsulated in \eqref{state2} and use it as the starting point. So we get
\begin{align}
\ln \Xi =-\frac{9 V}{\pi G } \int_0^\infty  \ln ( 1 - e^{ -
\beta \varepsilon } ) d \varepsilon= \frac{3 \pi }{2 G } \frac{V}{\beta}, \label{hololnx}
\end{align}
along with other thermodynamic properties
\begin{align}
E = - \frac{\partial }{\partial \beta}  \ln \Xi = \frac{3 \pi }{2 G } V T^2,  \label{energy} \\
S =   \ln \Xi  + \beta E =  \frac{3 \pi}{ G}  V T, \label{entropy}\\
P =T \frac{\partial \ln \Xi }{\partial V}  = \rho, \label{pressure}
 \end{align}
 with $V=\frac{4\pi}{3}R^3$. The gravitational energy of the system is $M=(1+3w)E=\frac{6 \pi }{ G } V T^2$ which naturally satisfies the Smarr formula $M=2TS$. Requiring $M=\frac{R}{2G}$, we deduce the Hawing temperature $T=\frac{1}{4\pi R}$. Substituting it into eq.\eqref{entropy}, we successfully obtain the exact Bekenstein--Hawking entropy
 \begin{align}
  S_{QG}= 4\pi G M^2= \frac{A}{4G}.
 \end{align}
Observing the logarithm of partition function \eqref{hololnx}, we soon realize that it is the same as that of a $1+1$ dimensional quantum system
\begin{align}
\ln \Xi =- \frac{ L_s}{\pi } \int_0^\infty  \ln ( 1 - e^{ -
\beta \varepsilon }  ) d\varepsilon =  \frac{\pi }{6 } \frac{ L_s}{ \beta}, \label{lnxstring}
\end{align}
 with size $L_s = 9 V/G$ \footnote{In \cite{xiao} the $1+1$ dimensional system was viewed as an effective string which has length $V/G$ and can vibrate in $9$ independent directions. The two descriptions are equivalent to each other at the leading order of the thermodynamics that we concern.}. Obviously this means that the black hole can be described as a $1+1$ dimensional quantum system at least from the thermodynamic viewpoint, despite the fact that it actually lives in a $3+1$ dimensional space-time. The derivation can be generalized to higher dimensions, and the necessity of $w=1$ for getting the exact Bekenstein--Hawking entropy has been emphasized in \cite{xiao}. In the Appendix, we further generalize the derivation above to general EoS $w$ in $D=d+1$ dimensions, where it contains an intriguing explaination about the equality between the exponent of the holographic entropy $S\sim M^\frac{D-2}{D-3}$ and the coefficient of the Smarr relation $M=\frac{D-2}{D-3}TS$.

Actually the thermodynamics \eqref{energy}-\eqref{pressure} has many interesting properties that may attach to quantum gravity. First, it has the EoS $w=P/\rho=1$. The fluid with $w=1$ is usually called stiff fluid  for that it is the most incompressible fluid permitted by relativistic causality. In contrast, black hole is also incompressible in some sense. If you want to accumulate more matter or entropy into it, the only way you can do is to increase its horizon size. Second, we can observe $S=\sqrt{6\pi} \sqrt{\frac{ EV}{G}}$ from these formulas. It has been shown that the expression $S\sim \sqrt{\frac{EV}{G}}$ is invariant under the T and S dualities, and it even keeps its form when we curl up some extra dimensions because of $\sqrt{\frac{E V_3 (L_c)^{D-4}}{G_{D-2}}}=\sqrt{\frac{E V_3}{G}}$ with $(L_c)^{D-4}$ the volume of the extra dimensions \cite{mathur}.

\section{Cosmological entropy and effective dimension} Now we have two typical kinds of constituents at hand, with respectively $w=1/3$ and $w=1$. For simplicity we start from a universe filled with a single kind of constitutes with a general EoS $w$. Consider a spatial-flat, homogenous and isotropic universe which is described by the Friedmann--Lema\^itre--Robertson--Walker (FLRW) metric and obeys the Friedmann equations
\begin{align}
 3\left( \frac{\dot{a}}{a}\right) ^{2}=8\pi G\rho, \label{4} \\
 \dot{\rho}+ 3\left( 1+w\right) \frac{\dot{a}}{a}\rho=0.\label{5}
 \end{align}
 The scaling behavior of the entropy inside particle horizon can be obtained in the following way \cite{fs}. From eq.\eqref{5} there is
\begin{align}
\rho=C_1 a(t)^{-3(1+w)}. \label{rho1}
\end{align}
 Substituting it into eq.\eqref{4}, there is
 \begin{align}
 a(t)=(\sqrt{\frac{8\pi G C_1}{3}} \frac{3(1+w)}{2}  t)^{\frac{2}{3(1+w)}},
 \end{align}
 which in turn gives
 \begin{align}
  \rho=\frac{3}{8\pi G} (\frac{2}{3(1+w)})^2 \frac{1}{t^2}. \label{rho2}
 \end{align}
Then the physical size of the particle horizon is
\begin{align}
R_{ph}=a(t)\int_{0}^{t}\frac{1}{a(t)}dt=\frac{3(1+w)}{1+3w}t. \label{18}
\end{align}
The energy inside the particle horizon is $E= \frac{4\pi}{3}R_{ph}^3  \rho=\frac{6(1+w)}{(1+3w)^3} \frac{t}{G}=\frac{2}{(1+3w)^2} \frac{R_{ph}}{G}$. The gravitational mass is thus
\begin{align}
M=(1+3w)E=\frac{2}{1+3w} \frac{R_{ph}}{G}. \label{21}
\end{align}
The cosmological expansion is an adiabatic process, so the entropy of the constituents in a co-moving volume $a(t)^3 s$ must be conserved. This leads to $s=C_2 a(t)^{-3} = C_3 t^{-\frac{2}{1+w}}$. Thus the entropy inside the particle horizon is
\begin{align}
S=\frac{4\pi}{3}R_{ph}^3 s = C_4 t^{\frac{1+3w}{1+w}}.
\end{align}
Using eq.\eqref{18}, the relation between the entropy $S$ and the particle horizon area $A=4\pi R_{ph}^2$ can be written as
\begin{align}
S \sim A^{\frac{1+3w}{2(1+w)}}. \label{ahorizon}
\end{align}
It shows the available entropy for an observer in the universe increases as the particle horizon expands, with different rates depending on $w$. This kind of cosmological entropy was first analyzed by Fischler and Susskind in applying holographic principle to cosmology \cite{fs}. The result is reliable since only the standard cosmological principles are used. It can also be written as $S\sim E^{\frac{1+3w}{1+w}}$ and generalized to the $D=d+1$ dimensional cases as $S \sim E^\frac{1+\frac{D-1}{D-3}w}{1+w}$ \cite{rama1}. Obviously, different kinds of constituents have different strategies of distributing energy into space.

Amazingly, even without knowing the microscopic physics of the constituents, eq.\eqref{ahorizon} reproduces the same scaling behaviors of the entropies as those we gained in last section. Concretely speaking, there are
\begin{align}
\begin{split}
S\sim A^{\frac{3}{4} } \ \ \text{for}\ \ \ w=\frac{1}{3};\\
 S\sim A  \  \ \text{for}\ \ \  w=1.
\end{split} \label{24}
\end{align}
Nevertheless, our calculation provides a clear statistical origin for eq.\eqref{24}. And below we need the corresponding thermodynamic properties to derive the temperature-time relation and the effective dimension of the universe.

We first discuss the temperature-time relation $T(t)$ of the universe and show that the relevant thermodynamic properties are consistent with the cosmological laws. For a universe filled with radiation, i.e., photons or other relativistic particles in conventional QFT, from eqs.\eqref{QFTe} and \eqref{QFTs} the entropy and energy density are respectively $s\sim T^3$ and $\rho \sim T^4$. Since the entropy in a co-moving volume $a^3 T^3$ is conserved, we get the familiar relation $T\sim a^{-1}\sim 1/\sqrt{t}$ for radiation dominated universe. And the corresponding energy density $\rho\sim T^4 \sim a^{-4}$ is consistent with the evolvement law \eqref{rho1} for $w=1/3$ universe. Similarly, for a universe filled with QG particles, from eqs.\eqref{energy} and \eqref{entropy} there are $s\sim T$ and $\rho \sim T^2$. Because $a^3 T$ is conserved now, we have $T\sim a^{-3}\sim t^{-1}$ which means the temperature changes more abrupt with time increases than that of the radiation case. And $\rho\sim T^2\sim a^{-6}$ is surely consistent with the evolvement law \eqref{rho1} for $w=1$ universe.

Then we come to the effective dimension of the universe. We have shown in last section that the QG system with $w=1$ is $1+1$ dimensional in essence. However, more formally we can measure the effective dimension of a system using the concept of thermodynamic dimension \cite{carlip1,carlip2}. The spirit of thermodynamic dimension is that partition function should depend on the dimension of phase space which certainly reveals the physically relevant dimensions at the quantum level. This can be translated to the temperature dependence of energy density. Thus, for a system consisting of massless particles, the effective dimension can be defined by $\rho \sim T^{D_{e}}$ or written as $D_{e}=\frac{d \ln\rho}{d \ln T}$. In the cosmological situation, due to eq.\eqref{rho2}, there is always $\rho\sim t^{-2}$. So we have $D_{e}=-2\frac{d \ln t }{d \ln T}$ showing that the effective dimension can be coded in the temperature-time relation $T(t)$. As what we expected, for the radiation dominated universe with $T(t)\sim 1/\sqrt{t}$ the effective dimension is $4$ and for QG particle dominated universe with $T(t) \sim 1/t$ the effective dimension is $2$, written clearly as
\begin{align}
\begin{split}
D_{e}=4 \ \ \text{for}\ \ \ w=\frac{1}{3};\\
 D_{e}=2 \  \ \text{for}\ \ \  w=1.
\end{split} \label{dimeff}
\end{align}
In the Appendix, we find that the effective dimension is $D_{e}=1+\frac{1}{w}$ for a $D$ dimensional universe filled with massless particles with EoS $w$.

\section{The running of effective dimension and entropy}
 The realistic universe with various kinds of constituents mixed together is far more complicated than that described above. When the constituents do not interact with each other, the energy density evolves like $\rho_i/\rho_j\sim a^{-3(w_i-w_j)}$, representing the overall trend of the universe to dilute the constituents with large $w$ and lower the average $w$, from $w=1$ to the conventional $w=1/3$ and $w=0$ and finally approaching to $w=-1$. On the other hand, above some characteristic temperatures and at high-energetic stages of the universe, the constituents actually interact strongly with each other and translate between. Only below the temperatures, they decouple from each other and evolve independently. In the following we provide a self-consistent scenario of evolvement of the universe, with emphasis on the running of effective dimension and entropy.

The first characteristic temperature we concern is roughly $k_B T= mc^2$ with $m$ the typical mass of nuclei. The interactions of standard model are responsible here to create massive particles. Obviously at this temperature the entropy density $s_m \sim \rho /m$ for massive particles and $s_r\sim T^3$ for radiation are of the same order, so the entropy can vary continuously in this process. Above the temperature the universe is radiation dominated, and below the temperature the radiation and matter start to decouple from each other, the universe evolves towards matter dominated. As for the effective dimension, the energy density for massive particles is $\rho=n mc^2+\frac{3}{2} n k_B T$, with the second term commonly omitted at $k_B T\ll mc^2$. Since the massive particles can freely move in $3$ directions, the number of spatial dimensions is obviously $3$. More formally the fact can be read from the energy equipartition term $\frac{D-1}{2}n k_B T$ \cite{carlip1}. Thus the effective dimension is fixed to be $3+1$ when the universe evolves from radiation dominated to matter dominated stage.

Another characteristic temperature is near the Planck scale, where the entropy density $s_r=T^3$ for radiation and $s_{QG}=\frac{1}{G}T$ for QG particles are of the same order. Thus, as the universe expands further and the temperature cools down, the holographic constituents may rapidly decay to or be diluted by other relativistic particles in standard model like photons. Accordingly the cosmological entropy evolves from a holographic form to the conventional form appropriate to radiation, and the effective dimension runs from $2$ to $4$. The transition begins from near Planck energy $10^{19} GeV$, and it is natural to assume it would finish to some extent at the Grand Unified Theory scale $10^{16} GeV$, which is also the scale where the gravitational interaction is believed to be separated from the other three fundamental interactions in nature. Then the remnants of the $w=1$ components can be further diluted to be inconspicuous by other components with lower $w$ in the following evolution of the universe. Generally, when the universe has expanded over $10l_p$, the concept of classical space and time can safely apply \footnote{The quantum gravitational corrections can be suppressed by a factor $(\frac{l_p}{L})^n$.}. So it justifies the using of the FLRW metric in the description of this process. The universe with EoS $w=1$ should be some analogue of black hole, and black hole can translate into radiation by Hawking evaporation, so maybe the same QG mechanism also plays its role here \footnote{The constituent of $w=1$ universe can be considered as a dense gas of black hole fluid from some perspective \cite{mathur,bf1,bf2}. The volume $V$ is taken as an arbitrary given volume in \cite{mathur} (in contrast we take it as the physical volume inside the particle horizon) so that it can contain many black holes. Here refers to another special property of the entropy formula $S\sim \sqrt\frac{EV}{G}$ that both a single black hole and a dense black hole fluid satisfy the form.}. Hawking evaporation has the ability to create all kinds of particles. Most of the standard model particles would be broken by the ultra-high-energetic photons, while those particles that do not interact with photons would retain and be explained as dark matter.

When tracing back to the time even earlier, it is basically not permitted to imagine a universe with $w>1$ which violets the relativistic causality. More likely, at this stage quantum effects are so strong that the classical geometric description of space-time is not applicable any more. The stage may be controlled by highly-excited strings or the so-called string-holes as suggested in \cite{rama1,venez}. Besides, requiring $D=2$ or $w=1$ naturally leads to a scale-invariant spectrum for cosmological perturbations even without inflation \cite{mukoh1,mukoh2,ac1,bf1,bf2}. Albeit this, our scenario puts no specific constraints on inflationary models and there were some attempts to combine inflation with $w=1$ universe \cite{mathur,bf3}.

It is interesting to note that various approaches to quantum gravity have suggested the dimensional reduction from $4$ to $2$ near Planck scale \cite{laus,amb,horava,carlip1,carlip2}. Our result shares a similar pattern with those of Ho\v{r}ava--Lifshitz gravity. Ho\v{r}ava suggested gravitational theories
 have the space-time anisotropy $\vec{x}\rightarrow b \vec{x}$, $t\rightarrow b^z t$ \cite{lifhor}. As $z$ flows from $z=3$ in the ultraviolet to $z=1$ in the infrared, the corresponding effective dimension changes from $4$ to $2$. In $D=d+1$ dimensions, the effective dimension for general $z$ is given by \cite{horava}
\begin{align}
D_s=1+\frac{D-1}{z}. \label{dimhora}
\end{align}
What happens in our context is that the number of quantum states $D(\varepsilon) d\varepsilon$ and thus the entropy are unchanged under the scaling transformation $\vec{x}\rightarrow b \vec{x}$, $ \varepsilon \rightarrow b^{-z} \varepsilon $, with $z=3$ for QG particles and $z=1$ for photons. In the Appendix we also provide the effective dimension for the general cases as
\begin{align}
D_{e}=1+\frac{D-1}{z}.  \label{dimus}
\end{align}
Though the expressions \eqref{dimhora} and \eqref{dimus} exactly match with each other, we may not naively take the whole frameworks to be conceptually equivalent. For example, in Ho\v{r}ava--Lifshitz gravity the anisotropic scaling of space-time is proposed to insure  power counting renormalizability and it modifies the Einstein-Hilbert action and the gravitational field equation. In contrast, we are searching for a non-trivial quantum matter satisfying the thermodynamics \eqref{thermo}. And our QG constituents determine the space-time geometry through the standard Friedmann equations with no modifications (or else we can not get the expected cosmological entropy with area scaling). Thus for now we regard this matching as mainly reflecting the universality of the fundamental QG theory.

\section{Discussions}
In conclusion, we have suggested that, when the universe evolves from a QG particle dominated universe with $w=1$ to a radiation dominated universe with $w=1/3$, the effective dimension runs from $2$ to $4$ and the cosmological entropy runs from $A$ to $A^{3/4}$. This may correspond to the phenomenon of dimensional reduction in the ultraviolet that has been found in various approaches to quantum gravity. The effective dimension afterwards is fixed to be $3+1$ even when the universe evolves to be matter dominated.

Here we stress again that the phenomenon of dimensional reduction doesn't necessarily mean a fluctuating space-time, it mainly reflects that some properties that we are accustomed to have to be greatly modified over some characteristic scale. For example, Ho\v{r}ava proposed an anisotropic scaling of space and time, motivated from critical phenomena of condensed matter physics \cite{horava}. In contrast, we modified the density of states for microscopic particles at high energy phase of the universe, with the motivation that there should be a holographic stage in early universe. The postulated density of states \eqref{state2} actually implies that the microscopic particles at this phase are quantized as if the space-time were $1+1$ dimensional, as is clear from eq.\eqref{lnxstring}. It seems the fundamental physics behind both black hole and early universe is $1+1$ dimensional \cite{xiao}, and this is worthy of further study.

Another interesting novelty of the work is that both the cosmological entropy formula $S=E^{ \frac{1+\frac{D-1}{D-3}w}{1+w} }$ and the effective dimension formula $D_{e}=1+\frac{D-1}{z}$ can be naturally derived from our microscopic setting. The same form of entropy was obtained from standard cosmological analysis in \cite{fs,rama1}, and the same form of effective dimension was obtained from Ho\v{r}ava--Lifshitz gravity in \cite{horava}. However, the two formulas were uncorrelated with each other in the previous literature, since they were derived in completely different contexts. Amazingly our work has suggested that the maximum entropy and the effective dimension are actually two aspects of the same phenomenon. Furthermore, all these physical concepts, i.e.,  maximum entropy, effective dimension, EoS $w$ and space-time anisotropy seem to be closely correlated, which is very interesting and has never been discovered before.

Our work also suggests to take seriously the $w=1$ stage of the early unverse. Fortunately the $w=1$ stage has already been conjectured and studied in cosmology for many years from a number of different physical motivations, with the properties like enhancing stochastic gravitational waves, dark matter abundance and baryon asymmetry \cite{zel,dut,mukoh3,chav2,nai}. The future observational evidence of the existence of such a stage would have profound implications for the understanding of quantum gravity \cite{Mureika:2011bv,Stojkovic:2014lha}.

\begin{acknowledgements}
I would like to thank X. Calmet for valuable discussions. I am grateful to MPS School of the University of Sussex for the research facilities and their hospitality during my visit. The work was supported by China Scholarship Council (No. 201908130079) and the Optical Engineering Key Subject Construction Project of Hebei University.
\end{acknowledgements}

\appendix

\section{The calculation for general cases}\label{app}

Here we aim to provide the results for general EoS $w$ in $D=d+1$ dimensions. We can generally consider the partition function
\begin{align}
\ln \Xi =g_1 V \int_0^\infty  \ln ( 1 - e^{ -
\beta \varepsilon } ) d \varepsilon^{  \frac{D-1}{ z }  }= g_2 V {\beta}^{- \frac{D-1}{ z } }, \label{general}
\end{align}
where $z=1$ corresponds to the familiar photons which we know the microscopic physics well and $z=D-1$ the conjectured QG particles. Notice the analysis is only applicable to massless particles, otherwise the parameters $m$ and $N$ have to be introduced. It follows from eq.\eqref{general} the thermodynamic properties $E=\frac{D-1}{ z }g_2 VT^{\frac{D-1}{ z }+1}$, $S=(\frac{D-1}{ z }+1)g_2 V T^{\frac{D-1}{ z }}$ and $w=P/\rho=\frac{z}{ D-1 }$. When $z$ continuously changes from $1$ to $D-1$, we get $w$ with value from $\frac{1}{D-1}$ to $1$.

For a general self-gravitational system there is $E\sim R^{D-3}$ and $P=w\rho \sim 1/R^2$, then one can check $dE+PdV$ can be written as $dM$ where $M \equiv (1+\frac{D-1}{D-3}w)E$. Thus the thermodynamic law can be written as $dS=\frac{1}{T}dM$. On the other hand, we can directly observe $M=\frac{1+\frac{D-1}{D-3}w}{1+w}TS$ from the above thermodynamic expressions, which gives $\frac{1}{T}=\frac{(1+\frac{D-1}{D-3}w)S}{ (1+w)M }$. So we have
$d S=\frac{(1+\frac{D-1}{D-3}w)S}{ (1+w)M }d M$ or written as $d \ln S = \frac{1+\frac{D-1}{D-3}w}{1+w} d \ln M$. It follows the result
\begin{align}
S \sim M^{ \frac{1+\frac{D-1}{D-3}w}{1+w} }.
\end{align}

The effective dimension for a $D=d+1$ dimensional universe can be evaluated by $D_{e}=-2 \frac{d \ln t }{d \ln T}$. Due to $T\sim t^{ -\frac{2}{1+\frac{ D-1}{z} } }$, we get
\begin{align}
D_{e}=1+\frac{D-1}{z}.
\end{align}
It can also be written as $D_{e}=1+ \frac{1}{w}$ by noting that $w=\frac{z}{D-1}$. Accordingly, for the special case $w=1$, we have the Smarr formula $M=\frac{D-2}{D-3}TS$, the holographic entropy $S \sim M^{ \frac{D-2}{D-3} }$ and the effective dimension $D_{e}=2$.

\end{document}